  %------------------begin definitions--------------------------------

  \def   \ni {\noindent}

  \def   \ssk {\vskip  5truept}
  \def   \sk  {\vskip 10truept}
  \def   \bsk {\vskip 15truept}

  \def   \newline {\hfil\break}

  \def\simlt{\lower.5ex\hbox{$\; \buildrel < \over \sim \;$}}
  \def\simgt{\lower.5ex\hbox{$\; \buildrel > \over \sim \;$}}
  
  \def\eg{{\rm e.g.$\,$}}
  \def\ie{{\rm i.e.$\,$}}
  \def\cf{{\rm cf.$\,$}}

  \def\l#1{\left#1}
  \def\r#1{\right#1}
  
  \def\apj{{\sl Ap.J.}}

  \def\mnras{{\sl MNRAS}}
  \def\nature{{\sl Nat}}

  \def\wtilde{\widetilde}  

  %------------end definitions-----------------------------------------

  \input psfig.sty
  \magnification=1000
  \hsize 5truein
  \vsize 8truein
  \font\abstract=cmr8
  
  \font\text=cmr10     
  \font\affiliation=cmssi10
  \font\author=cmss10
  
  \font\title=cmssbx10 scaled\magstep2

  \font\ita=cmti8
  \font\mma=cmr8
  \def\ref{\par\noindent\hangindent 15pt}
  \nopagenumbers
  \null
  \vskip 3.0truecm
  \baselineskip = 12pt
  
  {\title
  \ni Post-recombination gravity-generated contributions

  \ni to the cosmic microwave background anisotropies 

  \ni and cosmological parameter estimation.
  }
  \bsk \bsk
  {\author
  \ni Radek Stompor  \footnote{$^\dagger$}{\mma Current address: 
  The Center for Particle Astrophysics, 301 Le Conte Hall,
 
  \hskip 0pt \phantom{Current address:} University of California, Berkeley, CA 94720}
  }
  \bsk
  {\affiliation
   Institute of Astronomy, University of Cambridge, United Kingdom

   and Copernicus Astronomical Centre, Warszawa, Poland
  }

  \bsk
  %\cl {\it (Received )}
  \bsk
  \baselineskip = 9pt
  {\abstract
  \ni 
  Gravitational interaction of cosmic microwave background (CMB) photons
  with matter perturbations present along the line-of-sight to the
  surface of last scattering modifies the shape of the CMB anisotropy power spectrum. 
  Here I focus on (linear) integrated Sachs-Wolfe and (non-linear)
  gravitational lensing effects and discuss the detectability of
  the resulting distortions and their possible consequences for the CMB-based
  estimation of cosmological parameters. Specifically, I discuss
  if any of those effects may allow us to use CMB experiments to put
  independent constraints on the curvature of the universe  and the
  cosmological constant,
  i.e.$\,$ breaking the so-called {\ita geometrical degeneracy} in CMB
  parameter estimation discussed by Bond, Efstathiou \&\ Tegmark (1997)
  and Zaldarriaga, Spergel \&\ Seljak (1997).

  I address that issue using the Fisher matrix approach 
  and show that gravitational lensing of the CMB
  temperature and polarisation patterns might be detectable by 
  the Planck Surveyor satellite, leading to useful independent and
  precise constraints on the cosmological constant and spatial curvature.
  The integrated Sachs-Wolfe effect though bound to
  restrict those parameters only very weakly still may set constraints more
  stringent than those currently available.
  }
  \bsk
  \baselineskip = 12pt
  {\text                                                       %% beginning of font "text"
  \ni 1. INTRODUCTION.
  In standard inflationary scenarios of structure formation in the
  Universe the primordial perturbations generated during the inflation
  phase are amplified during the subsequent stages of the evolution,
  leading to the abundance of structures extending at the present from
  the scale well over $100h^{-1}$Mpc down to galactic scales.
  Byproducts of the matter perturbation evolution are the anisotropies
  in the photon background, which fills the Universe with a present day
  temperature of $T_0=2.726$K and energies in the
  microwave band. While these anisotropies are expected (and measured)
  to be very small, they are suitable for linear analysis and
  therefore provide a convenient tool for exploring the
  Universe. Furthermore, while they appear to be very sensitive to the evolution of
  the Universe as well as details of the perturbations present in the Universe
  CMB investigations offer an opportunity to constrain numerous important
  yet poorly till now known parameters. These include the parameters of the homogeneous
  FRW background as well as those describing
  statistical properties of the primordial perturbations.

  In any viable model the CMB photons stop interacting non-gravitationally
  with matter at very high ($z\simgt 50$) redshift (for definitness
  hereafter I will assume the standard thermal scenario (Peebles 1968)
  with the surface
  of last scattering (SLS) at $z\sim1100$ and neglect the possibility of
  inhomogeneous reionisation and existence of small-scale effects such
  as the Sunyaev-Zel'dovich effect). 
  Since then photons travel freely along
  the geodesics carrying on the image of the SLS.
   \midinsert
%   \vskip 0.5truecm
   \par\noindent
   \centerline{\psfig{figure=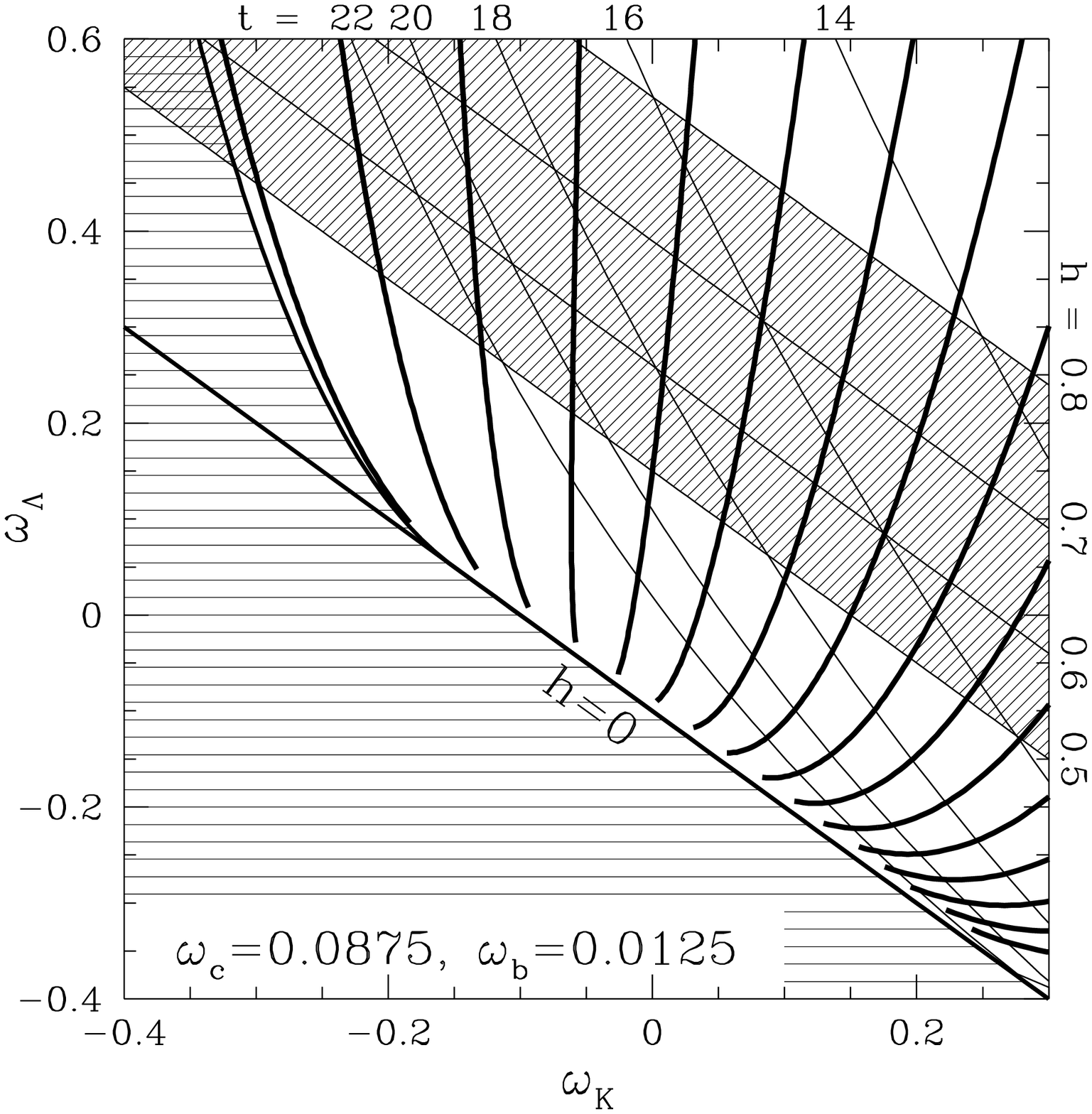,height=9.5truecm,width=9.5truecm}}
  
   \leftskip=1truecm
   \rightskip=1truecm
   \noindent
   {\abstract FIGURE 1. The thick solid lines show the locii of the
   cosmological models indistinguishable for CMB observations if 
   post-recombination contributions to observed anisotropies are
   neglected. Also, lines of the constant Hubble parameter
   (spanning the densely shaded area) and of constant age (in Gyr) are depicted
   with corresponding values as given at the edges of the figure.
   }
%   \vskip 0.5truecm
   \leftskip=0truecm
   \rightskip=0truecm
   \endinsert
  The high redshift of the SLS means that at that time the CMB photon distribution
  is not affected yet by either the cosmological % (time-independent)
  constant or the curvature of the Universe. Hence the models with very
  different values of those two parameters may produce identically
  perturbed universes at the time of last scattering. The angular scales
  at which those 3D structures are observed at the present are
  determined solely by one parameter: a comoving distance
  travelled by photons from SLS to the present denoted as $r_{SLS}\equiv
  r_{SLS}\l(\omega_b+\omega_c,\omega_h,\omega_\gamma,\omega_\nu,\omega_K,\omega_\Lambda\r)$~\footnote{$^\ddagger$}{\abstract I use physical densities 
  $\omega_i\equiv \Omega_ih^2,$ where $\Omega_i$ denotes, as usual, the density of the
  $i$th component in critical units, $h$ -- the Hubble parameter in units of
  $100$km/s/Mpc, and $i=b,c,h,\gamma,\nu,K,\Lambda$ stands for baryons, cold
  dark matter, hot dark matter, photons, massless neutrinos, the
  curvature and the cosmological constant respectively.}
  and usually referred to as the distance to the SLS.
  Consequently, any two models with different values of the
  cosmological constant and the curvature chosen to produce an identical distance to
  the SLS, and equal values of all the other parameters ($\omega_i$) may
  produce statistically indistinguishable patterns of
  observed CMB anisotropies if normalised appropriately (see Stompor
  \&\ Efstathiou 1998 (herafter SE98) for a correct normalisation procedure).
  That property of CMB anisotropies has been nicknamed a {\it
  geometrical degeneracy}.

  If the geometrical degeneracy were exact, then it would
  be a distance to the SLS which could be inferred from precise CMB 
  measurements, rather than values of the curvature and the cosmological constant
  separately. Those would remain almost unconstrained -- a serious setback for
  those hoping to use CMB for parameters estimation. Fortunately, it 
  appears not to be the case. 
  }
  \vskip 0.5cm
  {\text 
  \ni 2. POST-RECOMBINATION EPOCH.  
  As was noticed early on (\eg Sachs \&\ Wolfe 1967, Rees \&\ Sciama
  1968), though the Universe stays mostly neutral 
  after last scattering (by definition) the photons travelling through
  space are affected by the gravity of the growing matter
  inhomogeneities. The most important contribution is that on the largest
  angular scales where the anisotropy imprinted by the evolving
  potential wells may dominate the one generated at the time of last scattering.
  The effect, called the integrated Sachs-Wolfe (ISW) effect, is linear and
  may be produced either just after recombination (high-$z$ ISW), when
  radiation density still affects overall expansion or at
  low redshifts ($z\simlt10$) (low-$z$ or late ISW) with the curvature or/and the
  cosmological constant interfering with the pace of the expansion.
  As mentioned above, the amplitude of this effect, which is often
  dominant on the largest angular scales, is expected to decrease
  rapidly with angular scale, because gains and losses of the energy
  of photons crossing the potential wells of the perturbations with
  sizes much smaller than $r_{SLS}$ tend to average out very efficiently
  giving a negligible net effect.

  The remaining effects: the Rees-Sciama effect (or the non-linear ISW
  effect) and gravitational lensing are of second (and
  higher) order nature in a perturbative approach and therefore
  unavoidably small. Both were investigated in the past in
  considerable detail by many authors (see \eg Seljak 1996ab, SE98 and references therein)
 % (\eg Blanchard \&\ Schneider
 % 1987, Cole \&\ Efstathiou 1987, Sasaki 1987, Chodorowski 1991, 1992,
 % Atrio-Barandela \&\ Kashlinsky 1992, Seljak 1996ab), 
  who generally concluded
  that, of the two, it is lensing that might be of more importance in
  viable cosmologies. Hereafter I therefore discuss only the linear ISW
  and lensing effects. It is important to observe that though both
  effects are generated by gravity, it is only lensing that may
  distort a polarisation pattern. That property together with the
  fact that lensing-introduced distortions become larger on smaller
  angular scales implies that predominantly lensing is of importance for
  a CMB-based parameter estimation as demonstrated in the following.
  
  Both effects are displayed in Fig.$\,$2 and contrasted against
  the cosmic variance (full sky coverage assumed). What is presented
  in the figure are distortions of power spectra of the CMB
  anisotropies $C_\ell$ defined as a 
  variance of coefficients with a polar number $\ell$ of a multipole decomposition of the 
  two dimensional CMB maps into a series of spherical harmonics. In the realm
  of Gaussian theories the power spectra provide a complete
  statistical description of the CMB anisotropies.

   \midinsert
%   \vskip 0.5truecm
   \par\noindent
   \centerline{\psfig{figure=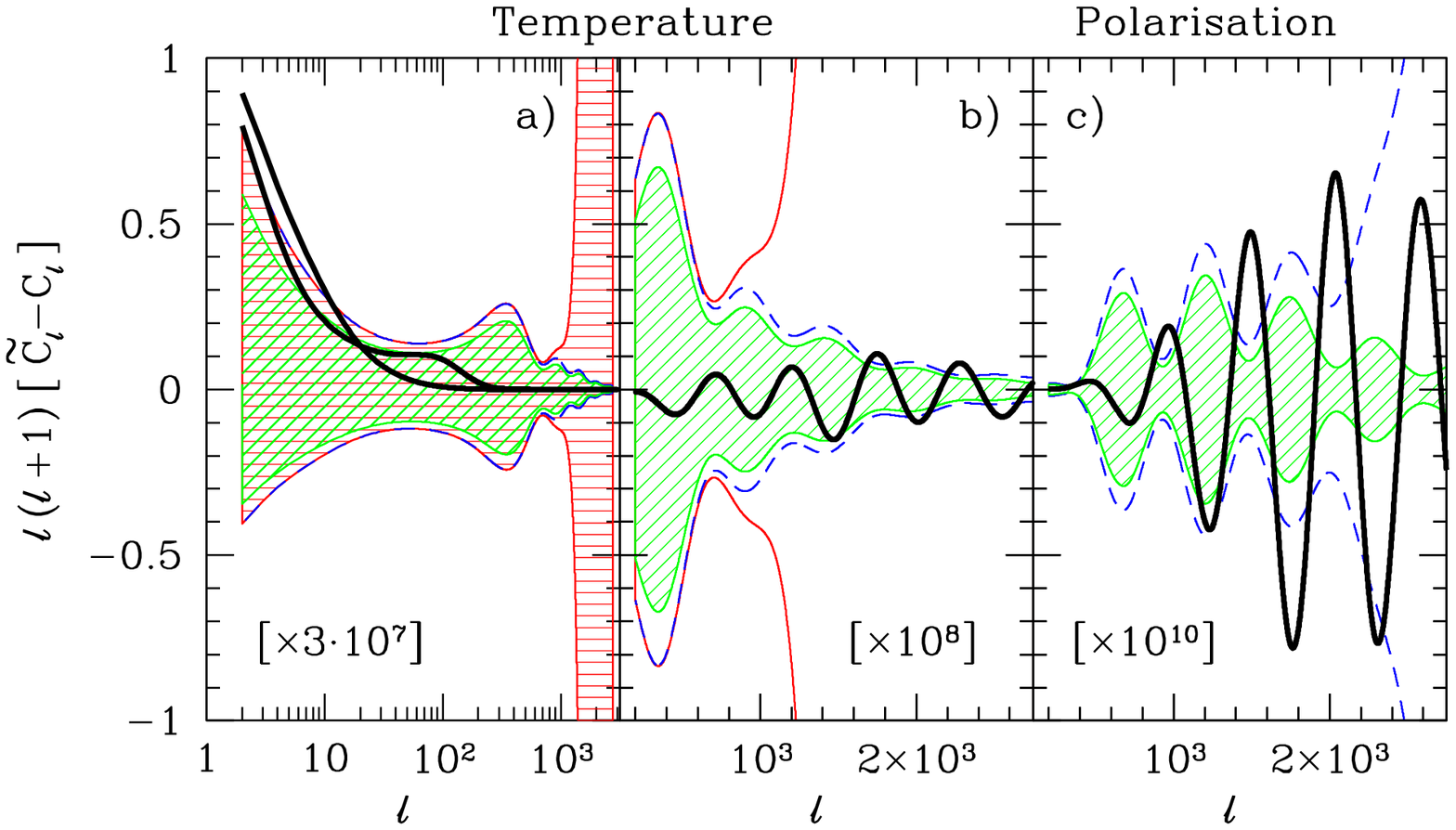,rheight=7.5truecm,width=13.0truecm}}
  
   \leftskip=1truecm
   \rightskip=1truecm
   \noindent
   {\abstract FIGURE 2. A comparison of the post-recombination contributions
   to CMB against the cosmic variance 1$\sigma$ uncertainty shown in each panel as
   a shaded area. The thick solid lines in the panel (a) correspond to low-$z$ 
   (a monotonically decreasing curve) and total ISW contributions. The
   horizontally shaded area shows an estimate of 1$\sigma$ errors for
   the MAP satellite. The thick solid lines in panels (b) and (c) show the
   gravitational lensing contribution to the temperature (b) and
   polarisation power spectra (c). The thin dashed lines in both
   panels mark an estimated 1$\sigma$ error for Planck, while the thin
   solid lines in (b) delineate errors for MAP. Assumed model
   parameters are:  $\omega_c=0.0875, \omega_b=0.0125, \omega_K=0.175,
   \omega_\Lambda=0.085$ with neither gravity waves nor reionization. $n_S=1$.
   }
%   \vskip 0.5truecm
   \leftskip=0truecm
   \rightskip=0truecm
   \endinsert

  The normalisation
  of the models follows the recent cluster abundance analyses giving a
  standard deviation of mass on scale of $8h^{-1}$Mpc equal to
  $\sigma_8\simeq0.52\Omega_0^{-0.6}$ (\eg Eke, Cole \&\ Frenk
  1996). %, Viana \&\ Liddle 1996). 
  To express the magnitude of discussed contributions 
  I employ here a $\chi^2$ statistics (per degree of freedom) defined 
  as (I neglect here possible cross-correlations between $C_\ell$ coefficients,
  see Wandelt, Hivon \&\ G\'orski 1998 for an exact treatment
  in a case of azimuthally symmetric sky cuts)
  $$ 
  \chi_\alpha^2\equiv{1\over \ell_{max}-1}\sum_{\ell=2}^{\ell_{max}}{\l(\wtilde
  Z_\ell^\alpha-Z_\ell^\alpha\r)^2\over \sigma_\alpha^2},\ \ \ \ \alpha=S \rm{\ or\ } BJ.
  $$
  Here $Z_\ell^\alpha$ is a function of $C_\ell$ solely and is
  considered to be a Gaussian variable. $\sigma_\alpha$ is the
  corresponding standard deviation. I consider here two possibilities.
  The first one is a standard high-$\ell$ approximation ($\alpha=S$) with
  $Z_\ell^S\equiv C_\ell$ and
  $\sigma_S^2\equiv2\l(2\ell+1\r)^{-1}f_{sky}^{-1}\l(C_\ell+w^{-1}b_\ell^{-2}\r)$.  
  The second choice ($\alpha=BJ$)
  follows that of Bond \&\ Jaffe (1998) and is supposed to account for 
  the non-gaussian character of $C_\ell$ and is therefore applicable on large
  angular scales (\ie at the low-$\ell$ tail). The appropriate definitions
  read in this case $Z_\ell^{BJ}\equiv\ln\l(C_\ell+w^{-1}b_\ell^{-2}\r)$
  and $\sigma_{BJ}^2\equiv 2\l[f_{sky}\l(2\ell+1\r)\r]^{-1}.$ Hereafter
  $b_\ell$ denotes an antenna beam pattern, and $w$ is a squared
  product of the noise level per pixel and the angular size of the pixels.
  The tilde over a quantity means that the post-recombination
  contribution has been taken into account. $\ell_{max}$ has been
  always chosen to maximize a corresponding $\chi^2$ value. The assumed specifications
  of the experimental setup follow those for the combined three best
  Planck channels. $f_{sky}$ is assumed to be 0.65.
  (See SE98 for a detailed description.)% and notations.) 

  For the particular cosmological model shown in fig.$\,$2, the
  corresponding $\chi^2$ values are $\chi^2_{BJ}\simeq 13.2$ (ISW) and
  $\chi^2_{BJ}\simeq 0.6$ (lensing) for the temperature power spectrum
  and $\chi^2_{BJ}\simeq 1.4$ for the polarisation one (lensing
  only). 
  It is therefore not only apparent that a strong detection of the ISW
  contribution is to be expected, but also that the lensing
  distortions might be detectable especially if a polarisation
  measurement is performed (\cf Fig.$\,$2c). Note that the major
  contribution to $\chi^2$ comes from the very low-$\ell$ modes in the
  ISW distortion case (\ie $\ell_{max}=2$) and the computed value can
  be improved upon neither by improving on the resolution or noise of
  an experiment. In the contrary, the major contribution in the lensing case
  comes from the highest accesible $\ell$-modes
  ($\ell_{max}\simeq 2650$ for a Planck-like observation) and therefore
  susceptible to further improvements.

  For lensing the ($\alpha=S$) choice of variables
  would give $\chi^2_S\simeq 0.6$ and $1.4$ for
  temperature and polarisation respectively, in good agreement
  with the previous numbers, showing that the use of these variables in
  lensing analyses is justified.

  \vskip 0.5cm
  {\text
   \ni 3. COSMOLOGICAL PARAMETERS ESTIMATION.
   While, as shown above, both ISW and lensing-generated distortions
   may be detectable for a Planck-type experiment, thus breaking
   the geometrical degeneracy, it is still to be determined
   to what precision this can be done.

   The fashionable way to investigate this problem is through the Fisher
   matrix analysis % Kendall \&\ Stuart 19.., 
   (\eg Tegmark, Taylor \&\ Heavens 1997). Here under the assumption
   of the nearly Gaussian character of the likelihood as a function of cosmological
   parameters, the expected uncertainties of the determination of
   cosmological parameters are estimated using the curvature matrix
   of the likelihood at its peak. 

   Involved computations require knowledge of the derivatives of the
   lensing contribution with respect to various cosmological
   parameters. In the case of gravitational lensing the smallness of these
   contributions  makes this a serious problem. Here I skip the
   description of the performed 
   calculations referring the reader to SE98 for a
   discussion. Derivatives of ISW distortions were estimated
   using a finite differencing scheme of the computed power spectra of
   the ISW effect alone.
   \sk
   {\abstract TABLE 1: The estimated uncertainties (1$\sigma$ standard deviations)
   of the determinations of the curvature $\l(\omega_K\r)$ and the
   cosmological constant $\l(\omega_\Lambda\r)$ based on the CMB
   observations for two cosmological models (no reionization nor
   gravity waves, $n_s=1$) with
   $\omega_c=0.0875,\omega_b=0.0125$ and $h=0.6$ and with either the lensing
   ($GL$) or ISW ($ISW$) contribution taken into account.
   I assume that temperature, polarisation and their cross correlation
   have been determined from the data.
   The results (multiplied by 100) are given for a Planck-like experiment.
  }
  \ssk
  {\hsize 5.0truein \settabs 6 \columns
  \hrule
  \+   \cr
  \hrule 
  \+   \cr
  \+ $\omega_K$  & $\omega_\Lambda$ & $\l[\delta \omega_K\r]_{GL}$ & $\l[\delta \omega_\Lambda\r]_{GL}$& $\l[\delta \omega_K\r]_{ISW}$ & $\l[\delta \omega_\Lambda\r]_{ISW}$ \cr 
  \+   \cr
  \+  0.0085 & 0.2515 & 0.1 & 1.2 & 0.8 & 10 \cr
%  \+   \cr
  \+  0.175 & 0.085   & 0.3 & 1.0 & 1.8 & 6.5\cr
  \+   \cr
  \hrule 
  }
  \sk    
  }
   Hereafter, I limit the parameter space to only two
   dimensions ($\omega_K,\omega_\Lambda$) with all other cosmological
   parameters fixed. A discussion of changes accompanying an increase
   in the
   dimensionality of the parameter space can be found in SE98. 
   Note, that % while $\partial Z^{BJ}_\ell/\partial C_\ell=\l(\sigma^S_\ell\r)^{-1}$
   the constraints set by the Fisher matrix analysis do not depend on
   the choice of $Z_\ell$ variables. A selection of the results is given in Table 1.

   \noindent
  \vskip 0.5cm
  {\text 
   \ni 4. CONCLUSIONS. The imprint of gravitational
   lensing on the power spectra of CMB anisotropies is likely to be detected
   by future high resolution and precision experiments including the instruments on board the Planck satellite.
   As a result, the non-linear gravitational lensing contribution 
   may have to be taken into account in the analysis of the future
   experimental data. One of the consequences of
   that fact, apart from the associated complications for the analysis
   itself, is rather fortunate. Lensing-generated distortions appear
   to contribute to breaking of the geometrical degeneracy, improving
   substantially on the constraints that can be set on 
   the curvature and the cosmological constant separately if only
   linear CMB anisotropy power spectra are considered.
   Though it is true in a case of both temperature and polarisation
   it is polarisation which appears to be much more sensitive to the 
   lensing-generated distortions.

   The forecasted errors in the determination
   of both of those parameters depend on the assumed normalisation of the
   primordial perturbations. However, keeping in mind the rather low
   normalisation inferred recently from the cluster abundance (and
   adopted for the work presented here), they are not expected to
   be underestimated too significantly.
   Hence high resolution and sensitivity CMB measurements
   on their own may
   provide an independent data set to be crosschecked for consistency
   with data from other sources in an attempt to verify theoretical assumptions
   underlying a choice of cosmological model parameter space under considerations.

   If lensing distortions are not unambiguously detected then the ISW effect
   may break the geometrical degeneracy of the linear CMB
   power spectra. However the inferred errors are almost an order of
   magnitude worse than those set with the help of lensing. Yet even those
   may seem attractive from the present-day perspective.

   The higher sensitivity of
   polarisation to the lensing effects makes its measurement not only
   desirable for the purpose of degeneracy breaking, but also as a test of
   the presence of nonlinear contributions to the temperature power spectrum.
  }
  \vskip 0.5cm 
  {\text
   \ni ACKNOWLEDGMENTS. I am grateful to George Efstathiou for
   collaboration on this project, and to Vince Eke for useful
   comments.

   The use of CMBFAST is acknowledged.
   This work was supported by UK PPARC grant and
   Polish Scientific Committee (KBN) grant No.$\,$2P03D00813.
  }
  \vskip 0.5truecm
  \ni {REFERENCES}
  \ssk
  \ref Bond J.R., Efstathiou G. \&\ Tegmark M., 1997, \mnras, {\bf 291}, L31
  \ref Bond J.R. \&\ Jaffe A.H., 1998, astro-ph/9808264
  \ref Eke V.R., Cole S. \&\ Frenk C.S., 1996, \mnras, 282, 263
  \ref Peebles, P.J.E., 1968, \apj, {\bf 153}, 1
  \ref Rees M.J. \&\ Sciama D.W., 1968, \nature, {\bf 233}, 395
  \ref Seljak U., 1996a, \apj, {\bf 460}, 549         % Rees-Sciama
  \ref Seljak U,  1996b, \apj, {\bf 463}, 1           % GL
  \ref Sachs R.K. \&\ Wolfe A.M., 1967, \apj, {\bf 147}, 73 
  \ref Stompor R. \&\ Efstathiou G.P., \mnras, in press, astro-ph/9805294 (SE98)
  \ref Tegmark M., Taylor A.N. \&\ Heavens A.F., 1997, \apj, {\bf 480}, 22
  \ref Wandelt B.D., Hivon, E. \&\ G\'orski K.M., 1998, astro-ph/9808292
  \ref Zaldarriaga M., Spergel D.N. \&\ Seljak U., 1997, \apj, {\bf 488}, 1  
                                                              %% end of the font "text"
  
  \end